\documentclass[12pt,twoside,onecolumn,a4paper,english]{IEEEtran2e}

\usepackage{a4wide}
\usepackage{amsmath}
\usepackage{amssymb}
\usepackage{fancybox}
\usepackage{floatfig,epsfig}
\usepackage{cite}
\usepackage{times}
\usepackage{babel}
\usepackage{algorithmic}
\usepackage{algorithm}
\newtheorem{example}{Example}[section]
\newtheorem{theorem}{Theorem}

\begin{document}

\title{\bf{Rate Adaptive Coded Digital Phase Modulation}}

\author{Yang Liu, Zihuai~Lin \\
 School of Electrical and Information Engineering, University of Sydney, NSW 2006, Australia\\ E-mail:
zihuai.lin@sydney.edu.au\\
}
\maketitle

\begin{abstract}
In this paper, rate adaptive coded Digital Phase Modulation (DPM) schemes based on punctured
ring convolutional codes are presented. We first present an upper
bound on symbol error probability for  punctured convolutional
coded Continuous Phase Modulation (CPM) over rings with Maximum
Likelihood Sequence Detection (MLSD). The bound is based on the
transfer function technique, which is modified and generalized to
punctured convolutional codes over rings.  
The novelty of this paper is the development of the analytical upper bound on the symbol error probability for the investigated system.
This work provides a systematic
way to analyze and design the rate adaptive ring convolutional coded CPM systems. The analysis method
is very general. It may be applied to any trellis based coding
schemes.
\end{abstract}

\pagenumbering{arabic} 

\section{Introduction}

Trellis coded Continuous Phase Modulation (CPM) is a bandwidth and
power efficient coded modulation scheme \cite{anderson}. It has
been studied in the past decades
\cite{anderson,ringcodedCPMyang,ringcodedCPMrimoldi}. One of the
advantages of trellis coded CPM is its constant envelope property,
which makes it a very suitable choice for data transmissions over
nonlinear and/or fading channels, \textit{e.g.,} satellite and
radio communication channels. This is because it can minimize the
distortion due to
non-linear amplification in the high power amplifiers. 
power amplifiers can be used, which can save $3-10$ dB
losses of non-linear distortion\cite{anderson}. 
higher power efficiency.
Because of the constant envelope property, these schemes are mostly suitable for 
fading channels, such as satellite and mobile radio channels.
We investigate a coded modulation scheme based on non-binary
trellis encoders, particularly, on ring convolutional encoders
\cite{ringCC89,ringCC90}.
The concept of convolutional codes over
rings is first introduced by Massey and Mittelholzer \textit{et
al} in \cite{ringCC89,ringCC90}. They show that when combined with
$M$-ary phase modulation, with the same number of states,
convolutional codes over the ring of integers modulo-$M$ have at
least as large a free Euclidean distance \cite{w&j} as the best
binary convolutional codes on the Galois field GF(2).

The applications of such schemes are for
data transmissions over
nonlinear and/or fading channels, \textit{e.g.,} satellite and
radio communication channels. 
schemes are usually considered as the most suitable choice.
Phase Shift Keying (PSK), which has been widely used
\textit{e.g.,} \cite{3GPP25814}, also has constant envelope. But
the discontinuous phase change between adjacent channel symbols
results in energy inefficiency. CPM has not only constant envelope
but also continues phase changes.
Compared with PSK, CPM has less
energy in its side lobe in their spectra, thus is more spectra
efficient.

Due to the excellent bandwidth efficiency, some non-constant
envelope schemes, 
\textit{e.g.,} $Q^2$PSK, SQAM \cite{SatelliteComm}, 
 can also be applicable to
satellite communications. 
In this case, however, the Traveling Wave Tube (TWT) type
amplifier, which is currently used in communication satellite
transponders and exhibit non-linear characteristics in both
amplitude and phase of the signal, must operate in the linear
region \cite{SatelliteComm}. This will increase the system
complexity and a significant loss of transmitter power.

The concept of convolutional codes over rings is firstly
introduced by Massey and Mittelholzer \textit{et al} in
\cite{ringCC89}. 
They show that when combined with
$M$-ary phase modulation, with the same number of states,
convolutional codes over the ring of integers modulo-$M$ have at
least as large a free Euclidean distance \cite{anderson} as the best
binary convolutional codes on the Galois field GF(2).  It is shown
\cite{ringcodedCPMyang,ringcodedCPMrimoldi} that compared to
trellis coded CPM  systems using binary codes of the same
complexity, ring convolutional coded CPM yield larger minimum
normalized squared Euclidean distance \cite{anderson}.

Most of the previous works in the area of ring convolutional coded
CPM, \textit{e.g.},
\cite{ringcodedCPMyang,ringcodedCPMrimoldi,ringcodedCPMlee}, focus
on searching the optimal codes in terms of the minimum Normalized
Squared Euclidean Distance (NSED). In \cite{ringcodedCPMMPEG},
ring convolutional coded CPFSK \cite{anderson}\footnote{CPFSK is a
subclass of CPM.} schemes are used for MPEG-4 image transmission.
There, Maximum \textit{a posterior} (MAP) sequence detection
(based on the Viterbi algorithm \cite{viterbialg}, 
is used for the recovery of the binary
bit streams from the MPEG-4 encoder. 
Serially concatenated CPM with a ring convolutional encoder as the
outer encoder is investigated in \cite{SCRingCPM}. In this paper,
we investigate rate flexible coded CPM over rings using Punctured
Ring Convolutional Codes (PRCC).
With adaptive coded CPM, the channel coding rate can be adapted to
the channel condition, therefore, the proposed scheme is more
robust compared with fixed rate coded CPM schemes.

For the un-interleaved PRCC/CPM scheme, an analytical upper bound on the bit error probability will be derived. This bound is based on
the union bound technique.

CPM schemes do not possess the uniform error property
\cite{viterbi}
 of geometrically uniform codes \cite{uniformcodes}.
Therefore, one can not assume the all zero sequence to be
transmitted during the analysis. In fact, CPM belongs to the class
of \textit{general codes} \cite{generalcodes}. Consequently, the
convergence analysis technique, \textit{e.g.},
\cite{exit,bitvssymbolinterleaver}, which was based on
geometrically uniform codes has to be modified to adapt to the
non-uniform error property of \textit{General codes}.

The main contributions of this work are: 1) the derivation of the analytical upper bound on the
error probability for the un-interleaved PRCC/CPM system; 2) the search of the optimal codes for serially concatenated punctured ring convolutional codes with CPM systems.

\section{Punctured Convolutional Coded CPM over Rings}

Let $Z_M$ denote the ring of integers modulo-$M$. For a
rate $1/n$ nonsystematic ring convolutional encoder, the generator
polynomial matrix is \cite{viterbi} 
$G_N(D)=
\begin{bmatrix}
G^0{(D)}\\
\vdots\\
G^{n-1}{(D)}\\
\end{bmatrix}$,
where $G^i(D)=g_0^i+g_1^iD+\cdots+g_m^iD^m$ for $i
\in \{0,\cdots,n-1\}$. The coefficients $g_j^i$ for $0 \leq j \leq m$
belong to the set $\{0,1,\cdots,M-1\}$. A high rate Punctured Ring
Convolutional Code (PRCC) can be obtained by puncturing a parent
$1/n$ ring convolutional code. The operation of puncturing some
coded symbols is implemented by using an $(n\times p)$ puncturing
matrix, $P_{mat}$, where $p$ is the puncturing period. Let $s$ be
the total number of transmitted symbols during a puncturing period
$p$, the coding rate of a PRCC is $r=p/s$.

Assume that the trellis of a PRCC is obtained by puncturing a rate
$1/n$ parent convolutional code. Let $L_s$ be the length of the
encoded source sequence.
The output of the PRCC encoder is then a sequence of $M$-ary
codewords, each of length $n$. 
length \footnote{Notice that the total length of the output
sequence of a PRCC eecoder is not related to the rate of the
underlying PRCC, but related to the rate of the parent code.}
of
the output sequence of the PRCC encoder is $n\cdot L_s$. The $i$th
codeword is $\textbf{u}_i$=($u_i^1,u_i^2,\cdots,u_i^{n}$),
$u_i^j\in\{0,\cdots,M-1\}$, $i\in\{1,\cdots,N\}$ and
$j\in\{1,\cdots,n\}$.

\section{Rate Adaptive Coded CPM using PRCC}\label{sec:ch6_adaptiveJSCC}

We now investigate a punctured trellis coded CPM over the ring of
integers modulo $M$.
The investigated CPM schemes are $M$-ary with
modulation index $h=1/M$. Both the punctured ring convolutional
encoder and the Continuous Phase Encoder
(CPE)\footnote{It was shown in \cite{Rimoldi} that CPM can be
described as the concatenation of a CPE and a memoryless
modulator (MM).} have the same algebraic structure. A block
diagram of the encoder is shown in Fig. \ref{fig:Ch6_integratedJSCCRingTCQCPM_2}. 
The generated
waveform from the CPM modulator is transmitted over the channel.
The channel is assumed to be the AWGN channel. 
\begin{figure}[htb]
\vspace{30ex}
\begin{picture}(-59,-40)(-60,-125)
\linethickness{1.0pt}
\setlength{\unitlength}{1.0mm}

\put(-12,-20){\vector(1,0){10}}
\put(-7,-15){\makebox(0,0){\footnotesize $\odot$}}

\put(-7,-20){\line(0,1){4}} \put(-7,-14){\line(0,1){5}}
\put(-7,-25){\makebox(0,0){\footnotesize $\odot$}}
\put(-9,-25){\makebox(0,0){\tiny $3$}}
\put(-7,-20){\line(0,-1){4}} \put(-7,-26){\line(0,-1){5}}
\put(-7,-9){\vector(1,0){18}}

 \put(13,-9){\vector(1,0){18}}
\put(33,-9){\vector(1,0){5}}

\put(-7,-31){\vector(1,0){37}} \put(32,-20){\vector(0,-1){10}}
\put(33,-31){\vector(1,0){5}}
\put(32,-31){\makebox(0,0){\footnotesize $\oplus$}}

 \put(-9,-15){\makebox(0,0){\tiny
$2$}} \put(-2,-23){\framebox(6,6){D}}\put(4,-20){\vector(1,0){15}}
\put(12,-20){\line(0,1){4}} \put(10,-15){\makebox(0,0){\tiny $2$}}
\put(12,-15){\makebox(0,0){\footnotesize $\odot$}}
\put(12,-14){\vector(0,1){4}}
\put(12,-9){\makebox(0,0){\footnotesize $\oplus$}}

\put(19,-23){\framebox(6,6){D}}\put(25,-20){\line(1,0){7}}
\put(32,-20){\vector(0,1){10}}
\put(32,-9){\makebox(0,0){\footnotesize $\oplus$}}

\put(45,-20){\vector(-4,3){5}}
\put(45,-20){\vector(1,0){2}}
\put(53,-20){\vector(1,0){9}}
\put(47,-23){\framebox(6,6){\footnotesize Punc.}}

\put(58,-20){\line(0,1){10}}\put(58,-10){\vector(1,0){33}}
\put(89,-7){\makebox(0,0){\footnotesize$\mu_{n}$}}


\put(65,-20){\circle{4}}\put(65,-20){\makebox(0,0){$+$}}
\put(67,-20){\vector(1,0){6}}\put(73,-23){\framebox(6,6){D}}
\put(79,-20){\vector(1,0){12}}\put(89,-17){\makebox(0,0){\footnotesize$v_{n}$}}
\put(91,-23){\framebox(8,16){MM}} \put(99,-15){\vector(1,0){10}}
\put(104,-12){\makebox(0,0){\footnotesize$s(t,\boldsymbol{\mu})$}}

\put(84,-20){\line(0,-1){10}}\put(65,-30){\line(1,0){19}}\put(65,-30){\vector(0,1){8}}
\put(65,-15){\makebox(0,0){\footnotesize{mod P}}}
\put(-11,-43){\dashbox(97,45)} \put(33,-2){\makebox(0,0){\small Concatenated punctured ring convolutional encoder with CPE}}
\end{picture}
 \caption{System model for the punctured ring convolutional coded CPM over rings, the generator polynomial matrix of the encoder is
 $G(D)=[2+2D+D^2;
3+D^2]$.}
\label{fig:Ch6_integratedJSCCRingTCQCPM_2}
\end{figure}

The overall channel encoder of the concatenated PRCC/CPM systems
is the combined PRCC encoder and the CPE. The CPE takes one output
symbol from the PRCC as an input and generates one vector which is
used by the memoryless modulator of CPM to generate one
channel transmission waveform. 
%
Decoding of the combined
system can be based on the trellis for trellis coded CPM over
rings. A system diagram for the concatenated PRCC/CPM without interleaver is given by Fig. \ref{fig:Ch7_PRTCQ/PRCCCPM}.

\begin{figure}[htb]
\vspace{20ex}
\begin{picture}(100,19)(-19,-9)
\linethickness{1.0pt} \setlength{\unitlength}{1.0mm}
\put(-2,23){\makebox(0,0){$\{{B}_k\}$}}


\put(0,20){\vector(1,0){7}} \put(7,16){\framebox(10,8){\small
ENC.}} \put(17,20){\vector(1,0){12}}
\put(29,16){\framebox(10,8){\small $P_{mat}$}}

\put(23,27){\makebox(0,0){$\overbrace{~~~~~~~~~~~~~~~~~~~~~~~~~~}$}}
\put(23,30){\makebox(0,0){\footnotesize PRCC encoder}}

\put(39,20){\vector(1,0){22}}

\put(50,23){\makebox(0,0){$\{\textbf{Y}_k\}/\{\boldsymbol{\mu}_k\}$}}
\put(62,16){\framebox(10,8){\small CPE}}
\put(72,20){\vector(1,0){11}}\put(77,23){\makebox(0,0){$\{\boldsymbol{\nu}_k\}$}}
\put(83,16){\framebox(10,8){\small MM}}
\put(77,30){\makebox(0,0){\footnotesize CPM modulator}}
\put(77,27){\makebox(0,0){$\overbrace{~~~~~~~~~~~~~~~~~~~~~~~~~~}$}}
\put(93,20){\line(1,0){12}}
\put(100,23){\makebox(0,0){$s(t,\boldsymbol{\mu})$}}


\put(105,20){\vector(0,-1){8}}
\put(95,2){\framebox(20,10){\footnotesize Channel}}
\put(105,2){\line(0,-1){8}}

\put(5,-6){\vector(-1,0){5}} \put(-2,-3){\makebox(0,0){$\{\hat
{{B}}_k\}$}} \put(5,-11){\framebox(25,10){\footnotesize Hard
Decod.}}\put(100,-9){\makebox(0,0){$r(t,\boldsymbol{\mu})$}}
\put(50,-6){\vector(-1,0){20}}\put(40,-4){\makebox(0,0){\footnotesize
$\{Pr({{B}}_k|\textbf{r})\}$}}
\put(60,5){\makebox(0,0){\footnotesize Joint Trellis}}
\put(60,2){\makebox(0,0){$\overbrace{~~~~~~~~~~~~~~~~~~~~~~~~~~}$}}
\put(50,-11){\framebox(20,10){\small APP Decod.}} 

\put(78,-11){\framebox(12,10){\small Demod.}}
\put(78,-6){\vector(-1,0){8}}
\put(75,-4){\makebox(0,0){$\textbf{r}$}}

\put(105,-6){\vector(-1,0){15}} 
\end{picture}
\vspace{10ex}
 \caption{System model for a concatenated PRCC/CPM system over the channel.}
\label{fig:Ch7_PRTCQ/PRCCCPM}
\end{figure}


We now study a symbol-by-symbol \textit{a posteriori} Probability
(APP) (or BCJR) decoding algorithm \cite{BCJR} for un-interleaved concatenated
PRCC/CPM over the ring of integers modulo-$M$. It is different
from all systems in the literature so far, \textit{e.g.},
 \cite{ringcodedCPMyang,ringcodedCPMrimoldi}, 
 etc.
For PRCC/CPM, the decoding is 
more complicated than that for convolutional coded CPM over rings
without puncturing \cite{ringcodedCPMyang,ringcodedCPMrimoldi}.
The reason is that the trellis structure varies for different
trellis sections.

For example, consider a combined punctured ring convolutional
encoder over $Z_4$ with $P_{mat}=[1 1 0;1 0 1]$ with a quaternary
1REC\footnote{$L$REC means the frequency pulse of the CPM scheme
is rectangular with pulse length $L$ symbol duration, see
\cite{anderson} for details.} CPM with modulation index $h=1/4$.
If at time $k$, there is no puncturing, (which corresponds to the
first column of the puncturing matrix $P_{mat}$), the decoder
employing the APP algorithm will determine the transition branch
metric based on the two received symbols or waveforms
corresponding to the two non-punctured output symbols of the ring
convolutional encoder. At time $k+1$, the decoder will determine
the metric based on the received symbol corresponding to the
non-punctured upper output symbol of the ring encoder. Since the
state of the CPE depends on the branch output symbols, the state
transitions will be different for different trellis sections.

It is possible to fix the trellis structure for decoding. 
The decoder can wait until receiving two symbols to decode even
for trellis sections with puncturing, that is, the decoding is
based on two trellis sections. 
The problem, however, is that the concatenated symbols of the two
trellis sections are not the output from the ring convolutional
encoder with one input symbol. They correspond to two consecutive
input symbols of the ring convolutional encoder. It is then hard
to determine the state of the concatenated PRCC/CPM.

In this work, 
the decoding is based on a varying trellis structure. The state
${\boldsymbol{\sigma}}_j$ of the trellis at discrete time $j$ is
defined as
$({\boldsymbol{\sigma}}^{cc}_j,{\boldsymbol{\sigma}}^{cpm}_j)$,
where ${\boldsymbol{\sigma}}^{cc}_j$ and
${\boldsymbol{\sigma}}^{cpm}_j$ denote the state of the PRCC and
the state of CPE at discrete time $j$, respectively. For a
concatenated PRCC/CPM system with a punctured ring convolutional
encoder over the ring of integers modulo $M$ having $m$ memory
elements, and an $M$-ary CPM scheme with a rational and
irreducible modulation index $h={\cal K}/P$, the total number of
states is
$M^m{\cdot}P{\cdot}M^{(L-1)}$. 
The state transition
${\boldsymbol{\sigma}}_j\rightarrow{\boldsymbol{\sigma}}_{j+1}$ is
determined by the input of the punctured ring convolutional
encoder. Associated with this transition is also the input symbol
$\mu\in\{0,\cdots,M-1\}$ of the CPE and the mean vector which is
obtained by letting the transmitted CPM waveform pass through a
bank of complex filters which are matched to the transmitted
signals.

The APP decoding algorithm \cite{BCJR} for an un-interleaved coded CPM system 
computes the APP $Pr(U_k=\mu|\textbf{r}_1^\ell)$ of an
input symbol $\mu$ of the CPE at symbol interval $k$ conditioned
on a sufficient statistic
$\textbf{r}_1^\ell=(\textbf{r}_1,\cdots,\textbf{r}_\ell)$ based on
channel observations $r(t,\boldsymbol{\mu})$, where $\ell$ is the
length of the input data sequence to the CPE.


In \cite{tcqcpmICC}, we developed an upper bound on the channel
distortion \footnote{The channel distortion is the signal
distortion caused by the noisy channel, further explained in
\cite{tcqcpmICC}.}
for a combined Trellis Coded Quantizer (TCQ)
with binary convolutional coded CPM system under MLSD.
In this work, we will derive an upper bound on the symbol error probability for the rate adaptive coded CPM system under MLSD. The bound
is based on the transfer function technique \cite{SER-bound},
which is modified and generalized to time variant trellis. This, in turn, is based upon the union bound. It is
shown in \cite{tcqcpmICC} that the developed analytical bounds are
consistent with the simulation results.
This method can also be
applied to the concatenated PRCC/CPM.
Let $E_s$ be the information
symbol energy and $N_0/2$ be the double sided power spectral
density of the additive white Gaussian noise. Let $m$ be the
number of memory elements of the  punctured ring convolutional
encoder and $\nu$ the transmission rate in bits per channel
symbol. The symbol error probability for a memoryless $M$-ary information source sequence of an un-interleaved PRCC/CPM system will follow the following theorem.
\begin{theorem}
Under the MLSD and the assumption that the source block is infinitely long, the symbol error probability for a concatenated PRCC/CPM system
with a discrete memoryless uniform digital source sequence, under the assumption that 
the upper bounds is based on the assumption that
the source block is infinitely long, so that the effects caused
by the error events across the boundary of the block can be
ignored.
In the appendix, it was shown that
 can be
upper bounded by 
\begin{eqnarray}\label{upboundsym2} 
P_s&{<}&Q\left(\sqrt{d_{min}^2\frac{E_sr}{\nu N_0}}\right)\exp\left(d_{min}^2\frac{E_sr}{2\nu N_0}\right){\cdot}
\nonumber\\
 &&
 \frac{{\partial}\Psi(\eta,\epsilon,\zeta)}{{\partial}\epsilon}\mid_{\eta=M^{-\nu},\epsilon=1,\zeta=e^{(-E_sr/{2\nu N_0})}},
\end{eqnarray}
where $d_{min}^2$ is the minimum NSED and $\eta$, $\epsilon$,
$\zeta$ are dummy variables \cite{SER-bound} and $r$ is the code rate of the trellis
encoder of the PRCC, $r=p/s$ as described in Section II. 
The average transfer function is 
\begin{eqnarray}
\Psi(\eta,\epsilon,\zeta)&=&M^{-m}\sum_{\kappa=1}^{M^m}\Psi(j,\kappa,\eta,\epsilon,\zeta)\nonumber\\
&=&M^{-m}\frac{1}{p}\sum_{j=0}^{p-1}\sum_{\kappa=1}^{M^m}\sum_{\iota}\sum_{\tau}\sum_d{\Theta_{j,s_\kappa,\iota,\tau,d}}\eta^{\iota}\epsilon^{\tau}\zeta^{d^2}
\end{eqnarray}
where $\Theta_{j,s_\kappa,\iota,\tau,d}$ is the number of error
events that start at time $j$ from state $s_\kappa$, and have NSED $d^2$,
length $\iota$ and total number of symbol errors caused by the
error event given by $\tau$. The $Q$ function is defined as
$Q(x)=(\sqrt{2\pi})^{-1}\int_{x}^{\infty}e^{-z^2/2}dz$. \hspace{25ex} $\square$
\end{theorem}

The proof of the above Theorem is given in Appendix A.

The transfer function $\Psi(j,\kappa,\eta,\epsilon,\zeta)$ can be obtained by using
a product state diagram \cite{Biglieri,productstate}. A product
state at time $j$ is defined as
$({\mathbf{\boldsymbol{\sigma}}}_j,{\mathbf{\boldsymbol{\hat\sigma}}}_j)$,
where ${\mathbf{\boldsymbol{\sigma}}}_j$ is a state of the encoder
of a coded CPM system and ${\mathbf{\boldsymbol{\hat\sigma}}}_j$
represents a state of the decoder. The transition
$({\mathbf{\boldsymbol{\sigma}}}_j,{\mathbf{\boldsymbol{\hat\sigma}}}_j){\rightarrow}
({\mathbf{\boldsymbol{\sigma}}}_{j+1},{\mathbf{\boldsymbol{\hat\sigma}}}_{j+1})$
is labeled with 
\begin{equation}\label{eq:labels}
 \sum_{\Delta{\tau}}\sum_{\Delta{d^2}}b(\Delta{\tau},\Delta{d^2})\eta{\epsilon^{\Delta{\tau}}}\zeta^{\Delta{d^2}},
\end{equation}
where $\Delta{\tau}$ and $\Delta{d^2}$ are the number of the
symbol errors and NSED, respectively.
$b(\Delta{\tau},\Delta{d^2})$ denotes the number of paths having
NSED $\Delta{d^2}$ and symbol errors $\Delta{\tau}$ for this
state transition.  

\begin{example}\label{ex:2}
For a punctured quaternary convolutional encoder with generator
polynomial $(1,D)$ combined with a quaternary $1$REC \footnote{$L$REC
means the frequency pulse of the CPM scheme is a Rectangular (REC)
with pulse length $L$ symbol duration \cite{cpm1,cpm2}.}
CPM system with $h=1/4$. 
The number of product states is $64$ and the product states range from $(0000)$ to $(3333)$.
\end{example}

For coded CPM over rings, 
the symbol error rate only depends on the output of
the ring convolutional encoder. In other words, it is independent
of the pair state of CPM
$(\mathbf{\boldsymbol{\sigma}}_j^{cpm},\mathbf{\boldsymbol{\hat\sigma}}_j^{cpm})$.
Furthermore, the NSED $d^2$ only depends on the difference of the CPM
states
$(\mathbf{\boldsymbol{\sigma}}_j^{cpm}-\mathbf{\boldsymbol{\hat\sigma}}_j^{cpm})$
\cite{anderson}. Therefore, the product state can be reduced. 
For full response CPM systems \cite{anderson}, the reduced product
state can be written as
$(\mathbf{\boldsymbol{\sigma}}_j^{cc},\mathbf{\boldsymbol{\hat\sigma}}_j^{cc},\omega_j)$,
where the difference phase state $\omega_j$ is given by

$\omega_j=\emph{R}_P\left\{(\mathbf{\boldsymbol{\sigma}}_j^{cpm}-\mathbf{\boldsymbol{\hat\sigma}}_j^{cpm})\right\}=\emph{R}_P\left\{\sum_{n=0}^{j-L}\gamma_n\right\}$
is the difference phase state, see (\ref{eq:d2}). 
The total number of product states for coded CPM with full
response CPM systems is $P{\cdot}M^{2m}$.


The product states of ring convolutional coded CPM can be divided
into initial states, transfer states and end states. A product
state is an initial state if an error event can start from it. A
product state is an end state if an error event can end in it. The
conditions for initial states and end states are
$\mathbf{\boldsymbol{\sigma}}_j^{cc}=\mathbf{\boldsymbol{\hat\sigma}}_j^{cc}$
and $\omega_j=0$. Other states are referred to as transfer states.

\begin{example}\label{ex:3}
For the system given in Example \ref{ex:2}, 
the total number
of the reduced product states is $64$. The initial states and the end states are the same, and they are
$(000)$, $(110)$, $(220)$, $(330)$. The remaining $60$ states are the transfer states. 
\end{example}

Let ${\mathbf{\boldsymbol{\xi}}}_{\kappa,j}$ represent the state
transitions from an initial state $s_\kappa$ to transfer states in
one step at time $j$. Let us denote by ${\mathbf{\boldsymbol{\Phi}}}_j$ the transitions
from transfer states to end states, and by
${\mathbf{\boldsymbol{\varpi}}}_j$ the transitions from transfer states to
transfer states in one step at time $j$. Let
${\mathbf{\boldsymbol{\chi}}}_{\kappa,j}$ represent the transitions
from an initial state $s_\kappa$ to end states in one step. The
transfer function can be calculated by
\cite{SER-bound,codedcpm}\vspace{-1ex}
\begin{equation}\vspace{-1ex}\label{eq:generatingfunction}
\Psi(j,\kappa,\eta,\epsilon,\zeta)=\mathbf{\boldsymbol{1}}{\cdot}\left({\mathbf{\boldsymbol{\Phi}}}_j({\mathbf{\boldsymbol{I}}}-{\mathbf{\boldsymbol{\varpi}}}_j)^{-1}{\mathbf{\boldsymbol{\xi}}}_{\kappa,j}+{\mathbf{\boldsymbol{\chi}}}_{\kappa,j}\right)
\end{equation}
where $\mathbf{\boldsymbol{1}}$ is an all one vector and
$\mathbf{\boldsymbol{I}}$ represents the identity matrix.

Equation (\ref{upboundsym2}) can be further expressed as
\begin{equation}\label{eq:asymptoticalser}\vspace{-1ex}
P_s {\leq} \sum_{d^2}\Theta_d{\cdot}Q(\sqrt{d^2\frac{E_sr}{\nu
N_0}}),
\end{equation}
where \vspace{1ex}
$\Theta_d=
\frac{M^{-m}}{p}\sum_{j=1}^p\sum_{\kappa=1}^{M^m}\sum_{\iota}\sum_{\tau}{\Theta_{j,s_\kappa,\iota,\tau,d}}{\cdot}{\tau}{\cdot}M^{-\nu{\cdot}\iota}$.

It can be seen from (\ref{eq:asymptoticalser}) that the minimum
NSED $d_{min}^2$ and $\Theta_{d_{min}}$ (the coefficient
of error events with $d_{min}^2$) dominate 
the asymptotical symbol error rate of the system. For computations of $\Theta_{j,s_\kappa,\iota,\tau,d}$ and $\Theta_d$, we refer the readers to \cite{LinZihuai_PhDthesis,tcqcpmICC}. 
%
For a concatenated PRCC/CPM, the NSED $d^2$ associated with an
error event $\Upsilon=\boldsymbol{\mu}- \boldsymbol{\mu}'$ can be
calculated as
\begin{equation}\label{eq:d2}\vspace{-0ex}\footnotesize
R{\cdot}\log_2M{\cdot}
(l-\frac{1}{T}\sum_{i=0}^{l-1}\int_{iT}^{(i+1)T}\cos[2{\pi}h\omega_i+4{\pi}h\sum_{j=i-L+1}^{i}\gamma_iq(t-jT)]dt)
\end{equation}
\begin{equation*}\label{eq:d2}\vspace{-0ex}
d^2=\frac{1}{2E_b}\int_{-\infty}^\infty[s(t,\textbf{U})-s(t,\hat{\textbf{U}})]^2dt
\end{equation*}.
\begin{equation}\label{eq:d2}
d^2=r{\cdot}\log_2M{\cdot}\left(\iota-\frac{1}{T}\sum_{i=0}^{\iota-1}\int_{iT}^{(i+1)T}\cos
\phi(t,{\boldsymbol{\gamma}})dt\right).
\end{equation}
Here $T$ is the symbol interval duration and
$\phi(t,{\boldsymbol{\gamma}})=[2{\pi}h\omega_i+4{\pi}h\sum_{j=i-L+1}^{i}\gamma_iq(t-jT)]$,
where
$\omega_j=\emph{R}_P\left\{({\boldsymbol{\sigma}}_j^{cpm}-{\boldsymbol{\hat\sigma}}_j^{cpm})\right\}=\emph{R}_P\left\{\sum_{n=0}^{j-L}\gamma_n\right\}$
is the difference phase state
and $r$ is the code rate of the trellis
encoder of the PRCC, $r=p/s$. 
$\emph{R}_x\{\cdot\}$ is the modulo $x$ operator and $q(\cdot)$ is
the phase response \cite{anderson}.

Largest $d_{min}^2$ is used as the design criterion for the
selection of the best puncture matrices of the investigated
un-interleaved PRCC/CPM systems.

\section{Simulation Results for Combined
PRCC/CPM}\label{sec:jscnumericalresults} Computer simulations
have been performed for rate adaptive
PRCC/CPM systems over AWGN channels. Three systems with puncture
matrices given by Table~1 are simulated. The PRCC encoders are
obtained by puncturing a rate $1/2$ $Z_4$ parent non-systematic
convolutional code with the generator polynomial matrix
$G(D)=[1+D+D^2;1+D^2]$ and $G=[1+ D^2; 1+2D+2D^2]$. For un-interleaved PRCC/CPM, the puncturing matrices are given by $P_{mat}^1$, which are obtained by an
exhaustive search from all of the possible puncturing patterns for
the one which gives the largest minimum NSED. The puncturing matrix is given in Octal form, for example, $(3,5)_o$ represents the puncture matrix $[0 1 1;1 0 1]$ where $0$ corresponds the position that the symbol is punctured.
The corresponding $\Theta_{d_{min}}$ are also given in the Table~1.
Also shown in Table~1 is the observation symbol intervals $N_B$ which is
needed to reach the upper bound on the minimum Euclidean distance \cite{anderson}.

\begin{table}[h!]
\begin{center}
{Table~1: Punct. matrices for the
investigated PRCC schemes.}\\
\vspace{2ex}\footnotesize

\begin{tabular}{|p{0.5cm}||p{1.0cm}|p{0.7cm}|p{0.7cm}||p{1.0cm}|p{0.7cm}|p{0.7cm}|}

\hline
Parent codes   & \multicolumn{3}{c||}{$G=[1+D+D^2;1+D^2]$}            &
\multicolumn{3}{c|}{$G=[1+D^2;1+2D+2D^2]$}   \\   
\hline
\hline Code Rate & $R=1/2$ $p=1$ &   $R={2}/{3}$ $p=4$ & $R=3/4$ $p=3$ & $R=1/2$ $p=1$ &   $R={2}/{3}$ $p=4$ & $R=3/4$ $p=3$   \\

\hline   $P^1_{mat}$ 
& $(1,1)_o$&$(15,13)_o$ &$(6,5)_o$  & $(1,1)_o$& $(16,15)_o$  &$(6,5)_o$ \\

\hline $d^2_{min}$ & 5.39 &  6.97 & 4.73  &6.54& 7.39 & $6.68$\\
\hline $\Theta_{d_{min}}$ & 0.0068  & 0.0396 & 0.0703 &0.0676&0.0779& 0.1758  \\
\hline $N_B$ & 11 &  13 & 6 &11&11&16 \\

  \hline
\end{tabular}
\end{center}
\end{table}

\begin{figure}[htb]
\begin{center}
\mbox{\epsfig{file=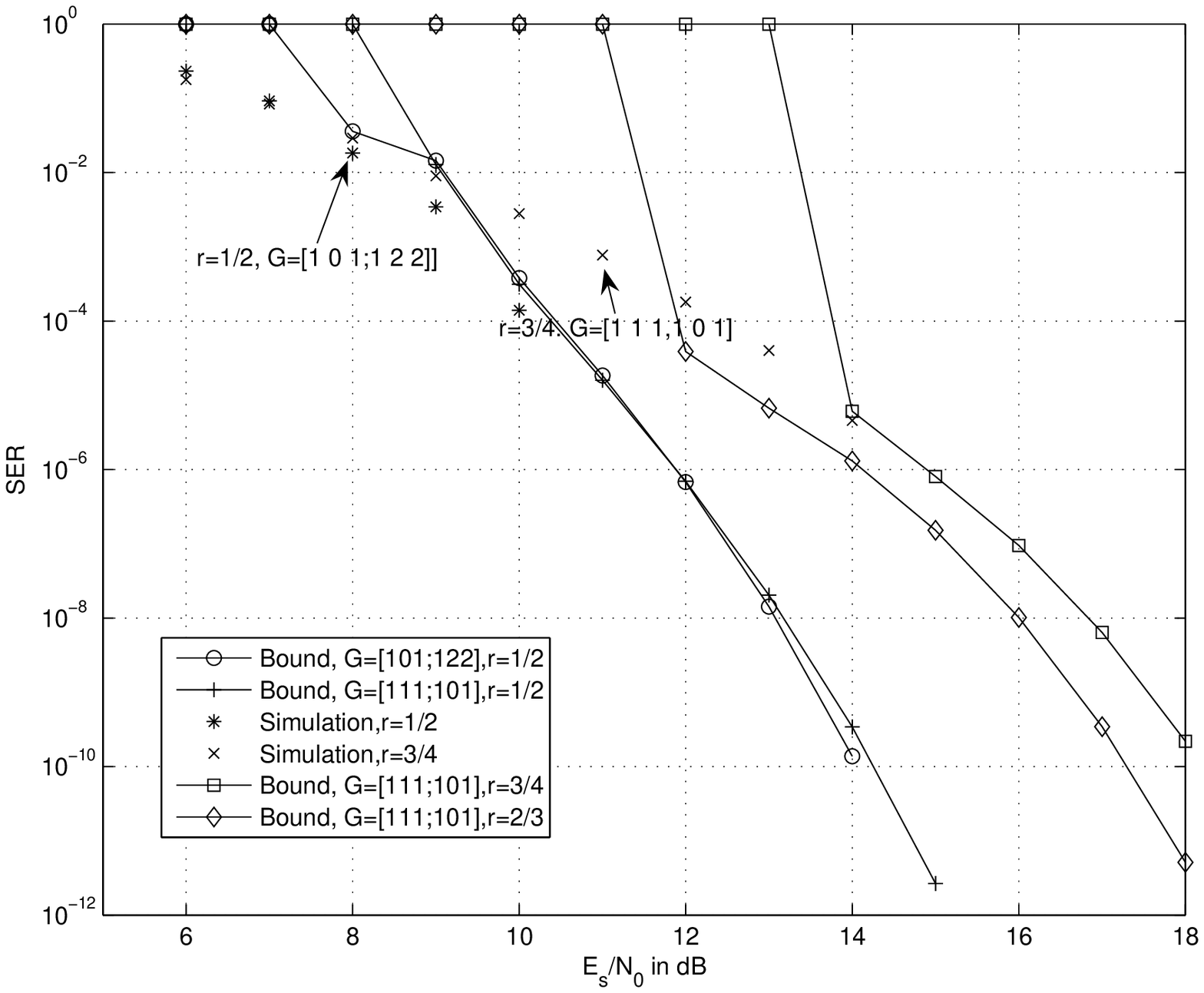,width=120mm}}
\vspace{20ex}
\caption{Analytical upper bound and some simulation results for un-interleaved PRCC/CPM with
$r=1/2,2/3,3/4$, $G=[1+D+D^2;1+D^2]$ and with $r=1/2$, $G=[1+D^2;1+2D+2D^2]$ .
} \label{fig:BERbound_PRCCCPM}
\vspace{6ex}
\end{center}
\end{figure}

For each channel SNR
value, $200$ source blocks with $5000$ symbols were used in the
simulation. Fig. \ref{fig:BERbound_PRCCCPM}  shows the upper bound and the simulation
results of BER performance for the investigated un-interleaved PRCC/CPM systems. It can be
seen that the simulation results agree with the upper bound especially when the channel SNR increases. It also shows that the
symbol error probability (SER) performance for parent codes is better than the punctured codes.  For example, for $G=[1+D+D^2;1+D^2]$, at SER of $10^{-4}$, 
a performance gain of roughly $\sim 2$ dB can be obtained
over the rate $3/4$ system.

\vspace{6ex}

\section{Summary}\label{sec:jscconclusion}

A rate adaptive coded modulation scheme based on concatenated PRCC/CPM
is investigated. The optimal puncturing matrices are presented. Compared with fixed rate convolutional coded CPM over ring, the channel coding rate can be adaptive to the channel condition, therefore it is more robust to the channel.

\section*{Appendix A}
The proof of the Theorem 1 is given below.
The ML sequence detector outputs the decoded source sequence according to
\begin{equation}
\hat{\textbf u}=\arg
\max_{\{{\textbf u}\}}\hspace{1ex}p(\textbf{r}|\textbf{u}).
\end{equation}
where $\textbf{u}$ is the source sequence, i.e., the input sequence of the punctured ring convolutional encoder.
Let us denote by $\varsigma_j$  the total symbol error caused by all error
events starting at a discrete time $j$. Let us take an arbitrary state,
say state $s$, $s\in{S}$, where $S$ is the state space. Let
$\Upsilon_{j,s,\iota,\tau,d}$ be the error event starting at time $j$
with initial state $s$, length $\iota$, and Normalized Squared
Euclidean Distance (NSED) $d^2$ \cite{anderson}. $\tau$ represents
the symbol error caused by the error event $\Upsilon_{j,s,\iota,\tau,d}$.
$\Upsilon_{j,s,\iota,\tau,d}$ is completely described by the start state
$s$ and the pair sequences
$(\mathbf{\boldsymbol{u}}_{s,\iota},\hat{\mathbf{\boldsymbol{u}}}_{s,\iota})$,
resulting in $\Upsilon_{j,s,\iota,\tau,d}$. Here
$\mathbf{\boldsymbol{u}}_{s,\iota}$ and
$\hat{\mathbf{\boldsymbol{u}}}_{s,\iota}$ are the reconstructed data
sequences.

For a concatenated PRCC/CPM, the NSED $d^2$ associated with an
error event $\Upsilon=\boldsymbol{\mu}- \boldsymbol{\mu}'$ can be
calculated as
\begin{equation}\label{eq:d2}
d^2=r{\cdot}\log_2M{\cdot}\left(\iota-\frac{1}{T}\sum_{i=0}^{\iota-1}\int_{iT}^{(i+1)T}\cos
\phi(t,{\boldsymbol{\gamma}})dt\right).
\end{equation}
Here $T$ is the symbol interval duration and
$\phi(t,{\boldsymbol{\gamma}})=[2{\pi}h\omega_i+4{\pi}h\sum_{j=i-L+1}^{i}\gamma_iq(t-jT)]$,
where
$\omega_j=\emph{R}_P\left\{({\boldsymbol{\sigma}}_j^{cpm}-{\boldsymbol{\hat\sigma}}_j^{cpm})\right\}=\emph{R}_P\left\{\sum_{n=0}^{j-L}\gamma_n\right\}$
is the difference phase state.
$\emph{R}_x\{\cdot\}$ is the modulo $x$ operator and $q(\cdot)$ is
the phase response \cite{anderson}.

The expected value of the symbol error rate, caused by error events
starting at time $j$, is given by
\begin{eqnarray}\label{eq:channeldist}
E[\varsigma_j]&=&
\sum_s\sum_\iota\sum_\tau\sum_d{\Theta_{j,s,\iota,\tau,d}}{\cdot}{\tau}{\cdot}Pr(\Upsilon_{j,s,\iota,\tau,d})\nonumber\\
&=&
\sum_s\sum_\iota\sum_\tau\sum_d{\Theta_{j,s,\iota,\tau,d}}{\cdot}{\tau}{\cdot}
Pr({\hat{\mathbf{\boldsymbol{U}}}}_j={\hat{\mathbf{\boldsymbol{u}}}}_{s,\iota}|{\mathbf{\boldsymbol{U}}}_j={\mathbf{\boldsymbol{u}}}_{s,\iota}){\cdot}Pr({\mathbf{\boldsymbol{U}}}_j={\mathbf{\boldsymbol{u}}}_{s,\iota}),
\end{eqnarray}
where ${\mathbf{\boldsymbol{U}}}_j$ and
${\hat{\mathbf{\boldsymbol{U}}}}_j$ are two random vectors, whose
outcome space are all possible
reconstructed signal sequences starting at time $j$.
$\Theta_{j,s,\iota,\tau,d}$ is the number of error events that start with
state $s$, and have NSED $d^2$, length $\iota$ and total error symbols
$\tau$. The expectation of (\ref{eq:channeldist}) is over all error
events starting at time $j$.

Suppose that the encoding rate of the convolutional encoder over
the ring $Z_M$ is $k/(k+1)$. Then there are $M^k$ branches
entering and leaving each state. For a discrete memoryless uniform
digital source, all sequences that start with a state $s$ and have
length $\iota$, are
equally probable. Thus, we can write 
\begin{equation}\label{eq:expectednofoevent}
Pr({\mathbf{\boldsymbol{U}}}_j={\mathbf{\boldsymbol{u}}}_{s,\iota})=Pr({\mathbf{\boldsymbol{\sigma}}}_j=s){\cdot}M^{-k{\cdot}\iota}.
\end{equation}

Let $\varphi$ be the set of all sequences
$\mathbf{\boldsymbol{U}}_{s,\iota}$ starting at time $j$ with state
$s$ having length $\iota$. Let us denote by $||\varphi||$ the cardinality of
the set $\varphi$. When $||\varphi||=2$, there are only two
sequences in the set $\varphi$, $\mathbf{\boldsymbol{u}}_{s,\iota}$
and $\hat{\mathbf{\boldsymbol{u}}}_{s,\iota}$. With MLSD, the decision
region of $\hat{\mathbf{\boldsymbol{u}}}_{s,\iota}$ is half of the
signal space and the error probability
$Pr(\hat{\mathbf{\boldsymbol{U}}}_j=\hat{\mathbf{\boldsymbol{u}}}_{s,\iota}|{\mathbf{\boldsymbol{U}}}_j={\mathbf{\boldsymbol{u}}}_{s,\iota})$
is exactly given by $Q(\sqrt{2d^2{E_b}/{2N_0}})$, where $E_b$ is the transmitted bit energy and 
$d^2$ is the NSED between the transmitted signals
$s(t,{\mathbf{\boldsymbol{u}}}_{s,\iota})$ and
$s(t,\hat{\mathbf{\boldsymbol{u}}}_{s,\iota})$. The complete signal
space squared distance between
$s(t,{\mathbf{\boldsymbol{u}}}_{s,\iota})$ and
$s(t,\hat{\mathbf{\boldsymbol{u}}}_{s,\iota})$ is given by
$2d^2{E_b}$. 
the transmitted bit energy
$E_b=E_sr/\nu$, where $\nu=\log_2M$. Therefore,
$Pr(\hat{\mathbf{\boldsymbol{U}}}_j=\hat{\mathbf{\boldsymbol{u}}}_{s,\iota}|{\mathbf{\boldsymbol{U}}}_j={\mathbf{\boldsymbol{u}}}_{s,\iota})=Q(\sqrt{d^2{E_sr}/{(\nu
N_0)}})$.
 When
$||\varphi||>2$, the decision region of
$\hat{\mathbf{\boldsymbol{u}}}_{s,\iota}$ is less than half of the
decision space. Therefore,
$Pr(\hat{\mathbf{\boldsymbol{U}}}_j=\hat{\mathbf{\boldsymbol{u}}}_{s,\iota}|{\mathbf{\boldsymbol{U}}}_j={\mathbf{\boldsymbol{u}}}_{s,\iota})<Q(\sqrt{d^2\frac{E_sr}{\nu
N_0}})$.
From above, we can conclude that
the conditional probability
$Pr(\hat{\mathbf{\boldsymbol{U}}}_j=\hat{\mathbf{\boldsymbol{u}}}_{s,\iota}|{\mathbf{\boldsymbol{U}}}_j={\mathbf{\boldsymbol{u}}}_{s,\iota})$
can be upper bounded by $Q(\sqrt{\frac{d^2{E_sr}}{{\nu N_0}}})$.

With the above result and (\ref{eq:expectednofoevent}), by considering
all error events and using the union bound technique \cite{viterbi},
$E[\varsigma_j]$
can be upper bounded by 
\begin{eqnarray}\label{eq:ej}\small
E[\varsigma_j]
\leq
\sum_{s}Pr({\mathbf{\boldsymbol{\sigma}}}_j=s)\sum_\iota\sum_\tau\sum_d{\Theta_{j,s,\iota,\tau,d}}{\cdot}{\tau}{\cdot}M^{-k{\cdot}\iota}
Q(\sqrt{d^2\frac{E_sr}{\nu N_0}}).
\end{eqnarray}

To further reduce the complexity of the computation of (\ref{eq:ej}), one need to note that for a coded CPM, some of the states are equivalent. Here ``equivalent" means that if the error events $\Upsilon_{\mathbf{\boldsymbol{\sigma}}}$ and
$\Upsilon_{\mathbf{\boldsymbol{\varrho}}}$, starting at time
$j$, generated by the same pair sequences
$(\mathbf{\boldsymbol{u}}_{\iota},\hat{\mathbf{\boldsymbol{u}}}_{\iota})$
starting at states $\mathbf{\boldsymbol{\sigma}}$
 and
$\mathbf{\boldsymbol{\varrho}}$, respectively, are identical.
Here ``identical" means that the error events have the same length,
the same NSED, and generate the same number of symbol errors.
In the following, we mathematically denote two equivalent states $\mathbf{\boldsymbol{\sigma}}$ and
$\mathbf{\boldsymbol{\varrho}}$ by
$\mathbf{\boldsymbol{\sigma}}\equiv{\mathbf{\boldsymbol{\varrho}}}$,

Based on the above statement, all the states of the punctured ring
convolutional coded CPM at time $j$, having the same state of the
convolutional encoder, $\mathbf{\boldsymbol{\sigma}}_j^{cc}$, are
equivalent. Among the $M^m{\cdot}P{\cdot}M^{(L-1)}$ states of the
coded CPM, there are only $M^m$ distinct states which are not
equivalent to each other. Consequently, (\ref{eq:ej}) can be
written
as 
\begin{eqnarray}\label{eq:ejeq1} \small
E[\varsigma_j]
\leq
\sum_{\kappa=1}^{M^m}Pr({\mathbf{\boldsymbol{\sigma}}}_j \equiv
s_\kappa)\sum_\iota\sum_\tau\sum_d{\Theta_{j,s_\kappa,\iota,\tau,d}}{\cdot}{\tau}{\cdot}M^{-k{\cdot}\iota} 
Q(\sqrt{d^2\frac{E_sr}{\nu N_0}}).
\end{eqnarray}
%

Let us denote by $L_T$ the length of the input sequence. Under the
assumption that the source sequence is infinitely long,
the symbol error probability can be upper bounded by 
\begin{eqnarray}\label{eq:ejeq2} 
P_s 
{\leq}
\frac{1}{L_T}\sum_{j=0}^{L_T-1}E[\varsigma_j]{\leq}\frac{1}{L_T}\sum_{j=0}^{L_T-1}\sum_{\kappa=1}^{M^m}Pr({\mathbf{\boldsymbol{\sigma}}}_j
\equiv s_\kappa){\cdot} 
\sum_\iota\sum_\tau\sum_d{\Theta_{j,s_\kappa,\iota,\tau,d}}{\cdot}
{\tau}{\cdot}M^{-k{\cdot}\iota}{\cdot}Q(\sqrt{d^2\frac{E_sr}{\nu
N_0}}).
\end{eqnarray}
For
an i.i.d. discrete memoryless source, the encoder can start at any
one of the $M^m$ trellis states. Further, since the
trellis is periodically time variant, $E[\varsigma_j]$ is periodically dependent on the time
$j$.
 The probability that a state
$\mathbf{\boldsymbol{\sigma}}_j$ of the coded CPM at time
$j$
is equivalent to one of the $M^m$ distinct states is 
\begin{equation}
Pr({\mathbf{\boldsymbol{\sigma}}}_j \equiv s_\kappa)=M^{-m},
\hspace{2ex} \forall{j}.
\end{equation}
Eq. (\ref{eq:ejeq2}) can then be expressed as\vspace{-1ex}
\begin{equation}\label{eq:ejeq3}
P_s {\leq}
\frac{1}{p}\sum_{j=0}^{p-1}M^{-m}\sum_{\kappa=1}^{M^m}\sum_\iota\sum_\tau\sum_{d}{\Theta_{j,s_\kappa,\iota,\tau,d}}{\cdot}{\tau}{\cdot}M^{-k{\cdot}\iota}{\cdot}Q(\sqrt{d^2\frac{E_sr}{\nu
N_0}}),
\end{equation}
where $p$ is the puncturing period as described in Section II, and $j=0$ corresponding the beginning of a puncturing period.

By using the inequality $Q(\sqrt{x+y})\leq{Q(\sqrt{x})e^{-y/2}}$
for $x,y>0$ \cite{anderson}, and letting
$d^2=d^2_{min}+(d^2-d^2_{min})$, the above inequality becomes
\begin{eqnarray}\label{eq:ejeq4}
P_s 
{\leq}
Q\left(\sqrt{d_{min}^2\frac{E_sr}{\nu
N_0}}\right)\exp\left(d_{min}^2\frac{E_sr}{2\nu
N_0}\right)M^{-m}\frac{1}{p}\sum_{j=0}^{p-1}\sum_{\kappa=1}^{M^m}
\sum_\iota\sum_\tau\sum_d{\Theta_{j,s_\kappa,\iota,\tau,d}}{\cdot}{\tau}{\cdot}2^{-k{\cdot}\iota}{\cdot}\exp({-d^2\frac{E_sr}{2\nu
N_0}}).
\end{eqnarray}

The right hand side of the above inequality can be computed by
means of the generating function. Let us denote by $\Psi(j,\kappa,\eta,\epsilon,\zeta)$
the generating function for one of the distinct equivalent states
$s_\kappa$, $\kappa=1,2,\ldots,M^m$. $\Psi(j,\kappa,\eta,\epsilon,\zeta)$ is then
given by 
\begin{equation}
\Psi(j,\kappa,\eta,\epsilon,\zeta)=\sum_\iota\sum_\tau\sum_d{\Theta_{j,s_\kappa,\iota,\tau,d}}\eta^\iota\epsilon^\tau\zeta^{d^2}
\end{equation}
where $\eta$, $\epsilon$, $\zeta$ are ``dummy" variables \cite{SER-bound}.
Using the above equation, Eq. (\ref{eq:ejeq4}) can be written as
\begin{eqnarray}\label{eq:ejeq5}
P_s
{\leq}
Q\left(\sqrt{d_{min}^2\frac{E_sr}{\nu N_0}}\right)\exp\left(d_{min}^2\frac{E_sr}{2\nu N_0}\right){\cdot}
\frac{{\partial}\Psi(\eta,\epsilon,\zeta)}{{\partial}\epsilon}\mid_{\eta=M^{-k},\epsilon=1,\zeta=e^{(-E_sr/{2\nu
N_0})}},
\end{eqnarray}
where the average generating function $\Psi(\eta,\epsilon,\zeta)$ is given
by 
\begin{equation}
\Psi(\eta,\epsilon,\zeta)=M^{-m}\frac{1}{p}\sum_{j=0}^{p-1}\sum_{\kappa=1}^{M^m}\Psi(j,\kappa,\eta,\epsilon,\zeta).
\end{equation}


\end{document}